1%
\documentclass[journal]{IEEEtran}

\ifCLASSINFOpdf
   \usepackage[pdftex]{graphicx}
\else
  \usepackage[dvips]{graphicx}
\fi
\usepackage{color}

\begin{document}
%
\title{Smart Charging Technologies for Portable Electronic Devices}
%
%
%

\author{Stefan Hild, Sean Leavey, Christian Gr\"af and Borja Sorazu
    
\thanks{S.~Hild, S.~Leavey and B.~Sorazu are with the School of Physics and Astronomy at the 
University of Glasgow, G12 8QQ, United Kingdom }
\thanks{C.~Gr\"af is with the Experimental Division, Albert-Einstein-Institut (Max-Planck-Institut f\"ur 
Gravitationsphysik), 30167 Hannover, Germany.}
\thanks{Manuscript received September 25, 2012;.}}

%
%

\markboth{Journal of \LaTeX\ Class Files,~Vol.~X, No.~X,}%
{Hild \MakeLowercase{\textit{et al.}}: Smart Charging Technologies for Portable Electronic Devices}
%



\maketitle

\begin{abstract}

In this article we describe our efforts of  extending demand-side control concepts to the application
in portable electronic devices, such as laptop computers, mobile phones and tablet computers.
As these devices feature built-in energy storage (in the form of batteries) and the ability to run complex
control routines, they are well-suited for the implementation of smart charging concepts. We
developed simple hardware 
and software based
prototypes of  smart charging controllers for  a laptop computer that steer the charging process depending on the 
frequency of the electricity grid and in case of the software implementation also based on the battery charge status. 
If similar techniques are incorporated into millions of devices in UK households,
this can contribute significantly to the stability of the electricity grid, help to mitigate the short-term power
production fluctuations from renewable energy sources and avoid the high cost of building and maintaining
conventional power plants as standby reserve.

\end{abstract}


%
\IEEEpeerreviewmaketitle

\section{Introduction}

In the recent past, many countries have strengthened their efforts to increase the fraction of electricity 
produced from renewable energy sources in order to reduce the greenhouse gas emissions 
from fossil fuel power plants.
For instance, in the Climate Change Act (2008), the UK government set out targets to cut CO$_2$
 equivalent emissions by 34\%
compared to 1990 levels by 2020 \cite{climatechangeact}. Additionally, in Scotland, the devolved government
has been considering plans ``to meet an equivalent of 100\,\% demand for electricity from renewable energy by 2020''
\cite{scottishrenewableplan}. The European Union also has a directive which legislates that the UK should
increase production of electricity from renewable energy sources by 15\,\% over the same period
\cite{eurenewableenergydirective}. Such policies are leading to an increase in the proportion of renewable
energy sources as a means of electricity supply.

The current electricity grid of the UK, similar to all other countries in the world, does not provide any 
significant storage capacity. The consequence is that each kilowatt-hour of electricity that we want to
 use somewhere within the country must be produced instantaneously at the very same moment, i.e. every 
 second our electricity production needs to be matched exactly to our electricity demand.

This requirement is probably the most demanding challenge towards the realization of a sustainable energy 
supply based on renewable energy sources such as photovoltaic and wind power, because 
their electricity output varies strongly with time, is often not reliably predictable and 
at most times differs greatly from the actual electricity demand. In particular, the lack of available
 renewable energy during prolonged calm and cloudy periods poses a major challenge. To provide enough 
 electricity on such days one can either resort to a large fleet of conventional power plants in standby mode, 
 ready to take over when the renewable energy sources cannot satisfy the demand, or build huge facilities 
 (such as pumped storage or batteries) for long duration electricity storage. Both of these approaches require
  costly hardware to be built and maintained on a scale similar to the existing fleet of conventional 
  power plants in the UK.
  
Smart grids and so-called demand control techniques \cite{Short2007}, \cite{Grid_smart}, \cite{Strbac2008}
provide a less costly and less hardware extensive alternative. This is achieved by relaxing the link between
electricity production and demand, based on introducing a mechanism such that the demand can be controlled to
match the variations in supply. This can be done by shifting less urgently required electricity consumption
out of times of high demand into periods of surplus energy production. For instance, if a person were to
connect their electric car or a laptop to a charger in the evening, it would probably not matter to the
individual exactly when during the night it is charged, as long as it is fully charged in the morning. The
development and introduction of smart charging technologies into as many areas of our daily life as possible
is the most promising solution for a reliable integration of renewable energy sources into the UK energy
portfolio. Ultimately, increasing the deployment of smart charging technology will allow us to reduce the
number of required standby power plants and to save the associated costs for construction and operation.
  
There is ongoing research on the use of demand control techniques in consumer hardware. Short et al.
\cite{Short2007} have demonstrated a demand controlled domestic refrigerator, showing that the technique
is useful in reducing the need for standby generation capacity when deployed on a large scale for both short
supply or demand transients and prolonged fluctuations caused by varying output from wind power. On a
commercial scale, the UK company Open Energi have recently started equipping the supermarket chain Sainsburys'
refrigeration portfolio with demand control technology, in the hope of saving 100,000 tons of CO$_2$
equivalent emissions over 10 years \cite{openenergi}.

There is also interest in incorporating the technology
into electric vehicles \cite{Amoroso} and plug-in hybrid electric vehicles \cite{Sortomme},\cite{Rotering}. Huang et al. \cite{Huang2010} describe an inductive power transfer mechanism for charging
electric cars which responds to grid frequency instabilities caused by an emulated wind turbine. The system was
shown to have improved frequency stability significantly, though the scope was limited to a single load on a
micro-grid. Work by Ifland et al. \cite{Ifland2011} outlines a theory considering the use of electric
vehicles as demand control devices, shaping their power load during charging to coincide with periods of high
supply. This is achieved through the use of power line communication and electricity market pricing mechanisms.
Similarly, Bashash et al. have developed a plug-in hybrid vehicle smart charger which optimises the charging 
process to the current electricity price and shapes load to increase battery longevity \cite{Bashash}.

In this article we extend the application of demand control concepts to portable electronic devices, such 
as laptop computers, mobile phones and tablet computers. We will describe the general principle of how to realize smart 
charging portable electronic devices in the next section. In 
Sections \ref{sec:CDS} and \ref{sec:uC} we discuss hardware prototypes of smart charging laptops, i.e. simple
external on-off controllers for laptop power supplies, similar to concepts that have been used for controlling
electric appliances \cite{DOE1},\cite{DOE2}.
 As we will discuss in Section \ref{sec:soft}, in a second step we also realised a  smart charging 
laptop that is entirely based on software, which controls the charging process by supervising the battery management control 
system.
 Future developments including the application of more complex charging 
algorithms are discussed in Section  \ref{sec:future}. Finally we give a conclusion  in Section \ref{sec:con}.

\section{The concept of smart charging of portable electronic devices}
\label{sec:principle}
\subsection{The principle}
Portable electronic devices, such as laptop computers, mobile phones and 
tablet computers\footnote{In 
the following we will always use a laptop computer as our example, but similar technologies
can also be applied to tablets, phones and other portable electronic devices.}
can easily be transformed to ``be smart'' in terms of energy handling.
First of all, without any additional hardware or cost they already feature built-in 
electricity storage (the battery). Additionally, these devices are inherently capable of 
performing complex algorithms, required for smart charging, as well as 
communicating with other devices, for instance via the Internet to receive or 
exchange information on the status of the electricity grid.
   
So, imagine the following scenario: in a few years from now we can all choose in the 
power management settings of our laptops not only options for \emph{long battery life} 
or \emph{high laptop performance}, but also an option which can be set to \emph{smart charging} in order
 to mitigate power fluctuations from time-varying renewable energy sources. This could be incentivized either 
 through social responsibility or by power companies which could reward customers for using the setting by
  monitoring energy consumption with in-home smart meters.

\begin{figure}[h]
\centering
\includegraphics[width=\columnwidth]{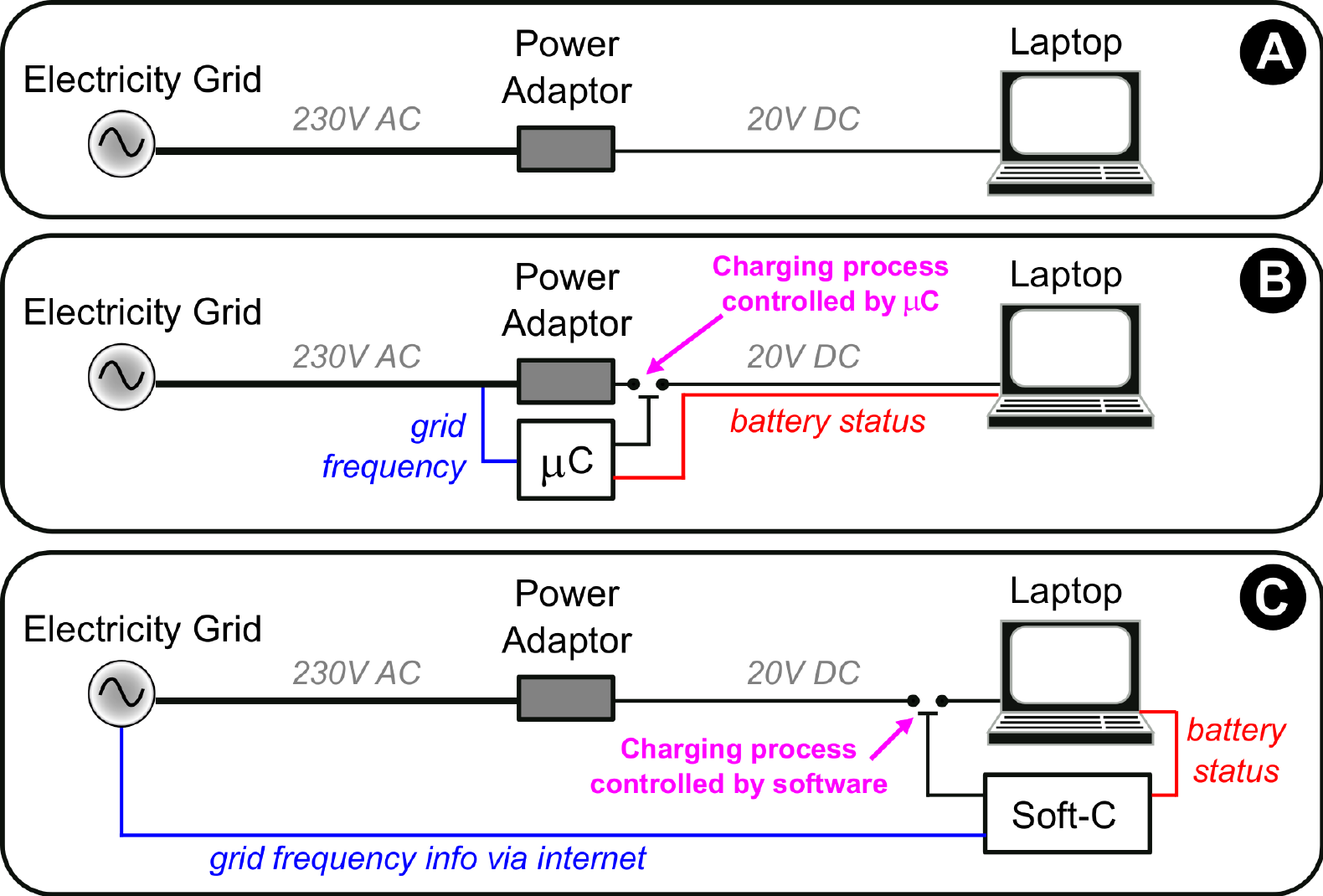}
\caption{Sketch illustrating the principle of smart charging laptops. \textbf{A)} 
a standard laptop. \textbf{B)} A microcontroller ($\mu$C) performs a local 
measurement of the grid frequency and receives the battery charge state information from
the laptop. On the basis of these two pieces of information an algorithm in the microcontroller 
controls the charging process. (Instead of the local frequency measurement
the grid frequency can also be communicated to the microcontroller by means
of a higher harmonic modulation of the grid frequency itself). \textbf{C)} The charging 
process is guided by a software controller (Soft-C) inside the laptop using the 
battery charge state information as well as the grid frequency information which could be 
received from a central server via the Internet.}\label{fig:concepts}
\end{figure}

How would this work in principle? First of all, the laptop would need to know whether there is a deficit or 
a surplus in the electricity grid. This can either be obtained from a central server via the Internet 
(or from a server controlled by the user's power company), 
or deduced from a local measurement of the grid frequency, itself an indicator of the electricity 
supply-to-demand ratio in the national grid \cite{Short2007}. A grid frequency above 50\,Hz indicates
a surplus of electricity in the grid, whereas a grid frequency below 50\,Hz indicates that the electricity
demand outweighs the electricity supply.\footnote{Please note that throughout this article we refer to 
the grid frequency as 50\,Hz. However, the same techniques described here can be easily transferred 
to electricity grids using a frequency of 60\,Hz.}  
 Another way to transfer the information of the electricity 
grid from a central location to individual consumers would be usage of high frequency amplitude modulations
of the grid voltage, similar to as is done to control night storage heating tariffs or the switching of street lamps.
Then depending on whether there is too much 
or too little electricity in the grid and also taking the current battery charge status and the 
user's habits into account, a controller could steer the charging of the laptop in a smart 
way.   Such a controller can either be implemented in hardware inside the power adaptor or the laptop 
itself, or it could simply be a software algorithm steering the charging of the battery. As we will 
discuss in more detail in Section \ref{sec:future}, when it comes to electronic devices,
different individuals have different personal user habits, and therefore personalized 
charging profiles are bound to vary from case to case. For instance, a particular user might insist upon a minimum level of charge that 
should always be available, while others may require a fully charged laptop at a particular 
time of day (such as before a morning train journey).

Figure \ref{fig:concepts} gives a simplified overview of different principles for smart charging of portable
electronic devices. The schematic drawing in the top frame of Figure \ref{fig:concepts}, denoted as A,
 illustrates the charging of a standard laptop. A power adaptor is connected
to the electricity grid and converts the 230\,V AC into about 15--20\,V DC (the exact value varies for different
devices and manufacturers).
The DC voltage is connected to the laptop and depending on the battery status the charging is switched on 
or off within the laptop itself.

The sketch in the middle frame  of Figure \ref{fig:concepts}, labeled B,  depicts a setup in which a simple external on-off controller
 is used
to control the DC power supply connection between the power adaptor and the laptop. In this 
example the controller unit is implemented in hardware, for instance by analog electronics or a microcontroller 
($\mu$C). Such a microcontroller can be used for measuring the grid frequency, to deduce the current supply to
demand ratio of the electricty grid (see Section \ref{sec:uC} for 
details of an example implementation). Ideally, the microcontroller has knowledge of 
the actual charge status of the laptop battery. This information, the grid frequency and the charge status 
of the laptop battery, can be sent to software running on the microcontroller to generate a signal, 
which can then be used to control the smart 
charging process, e.g. by toggling an 
electronic switch.

A cheaper alternative to the hardware implementation of a smart charging laptop can be realized purely
in software (see plot C of Figure \ref{fig:concepts}). As most portable electronic devices are already 
now connected to the Internet for a major fraction of their time in use (and it seems likely 
that in the future the fraction will only increase), it can be more efficient to  perform the grid frequency measurement centrally and 
send the information via the Internet to millions of devices. This option would not only save on costs related to hardware,
but also allow the implementation of a measurement setup providing a much higher precision than is feasible in the low-cost
setup included in every power adaptor or laptop. The grid status information received from a central server can then again
be combined with the battery charge status and the personal user preferences and used as input for an algorithm that controls
the charging process.


Obviously, the scenarios described above illustrate only two examples of a range of possible implementations.
Combinations of various elements of the concepts shown in Figure  \ref{fig:concepts} B and C are possible.
In Sections \ref{sec:CDS} and \ref{sec:uC} 
we present and discuss first hardware prototypes of smart charging laptops, while in Section \ref{sec:soft} we 
present a software based implementation of a smart laptop charger that we realized.

\subsection{Smart charging and battery health}
If the smart charging concept described above is applied to portable electronic devices
then the characteristics of the charge and discharge cycles may change slightly. When
a normal laptop computer is connected to its power supply it charges the battery 
continuously up to full charge and then trickles. In contrast when connecting a smart charging 
laptop system to the electricity grid it would usually not fully charge the battery in a single go,
but rather there might be some interruptions in the charging during periods in which the electricity demand
in the grid outweighs the available supply. In addition the integrated duration of being in trickle mode will be reduced.

Both of these differences originating from the application of smart charging would not have any serious 
negative effects on the health of Lithium batteries. First of all, studies have shown that there is no significant 
correlation between the depth of discharge (DOD) and the battery degradation, but only 
a correlation between total processed capacity of the battery and the battery
degradation \cite{Peterson2010}. As the smart charging concept would only change the 
DOD, not the total processed capacity of the battery, we do not expect any negative effects.
Moreover, the non-continuous charging when using a smart controller would probably lead to a
 reduced peak temperature during the charging process, which would actually reduce the battery
  degradation \cite{Bloom2001}.

%
%

\subsection{Benefits of smart charging of portable electronic devices}
Obviously, due to their significantly smaller capacity, smart charging of portable electronic 
devices has in the long term a lower potential of demand to supply balancing than other currently discussed approaches
such as smart charging of electric cars. However, in the short 
and medium term future the number of existing portable electronic devices is orders 
of magnitude larger than the number of existing electric cars. Also, as laptop computers 
and other portable electronic devices anyway always have a rechargeable battery, the smart charging
concept can be integrated in future hardware without any significant cost.

Let us assume that 2/3 of the UK population were to use a laptop (or similar device) and half of these 
laptops would be connected to a charger at a time. Then 20 million devices would be connected to 
the National Grid at any given time. If we further assume that these devices would charge on average with a 
power of 50\,W each, then if optimally deployed the smart charging concept for laptops 
alone would allow us to avoid building and operating on standby about one large power plant of 
the 1\,GW class  and save the associated costs. A more comprehensive  analysis of the large scale 
effect of our concept onto the electricity grid is currently under way and will be published in a separate 
article.

Moreover, this technology will allow each individual laptop user to save money, as it is expected 
that in future the price for electricity will not be constant, but vary depending on the actual demand 
and availability of electricity \cite{Strbac2008}. 

\begin{figure}[htb]
\centering
\includegraphics[width=\columnwidth]{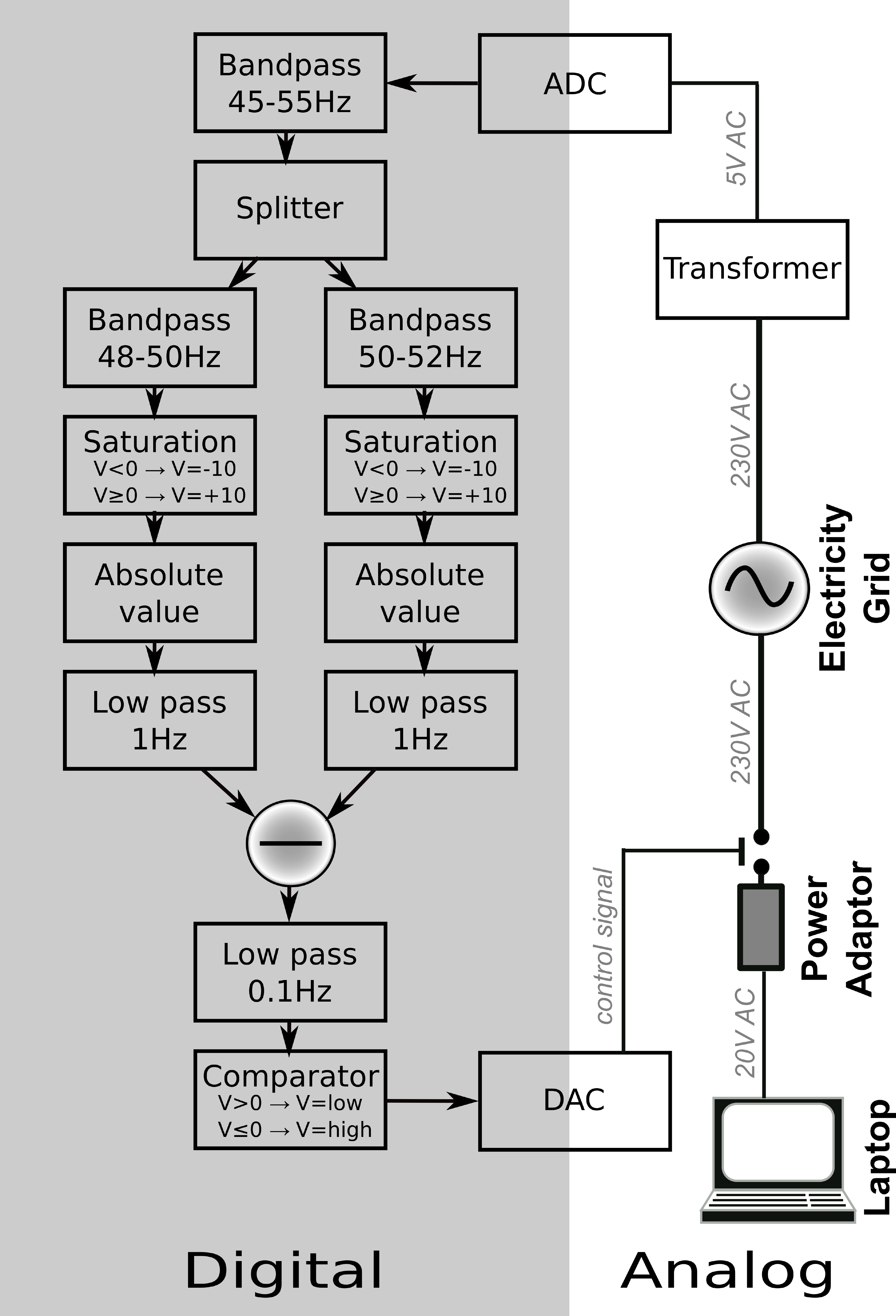}
\caption{Schematic of a simple external on-off implementation of a smart charging laptop. The grid 
frequency is analysed by a digital system using a differential bandpass technique (see a more detailed explanation in the text).
If the grid frequency is above a certain frequency (for instance 50.00\,Hz) then
the digital control system switches on the power to the AC adaptor of the 
laptop. If the grid frequency is below the set threshold, the charging of the 
laptop is suspended.  (Used abbreviations: ADC = analog-to-digital converter, DAC =
digital-to-analog converter.)} \label{fig:CDS_setup}
\end{figure}

\begin{figure*}[htb]
\centering
\includegraphics[width=0.95\textwidth]{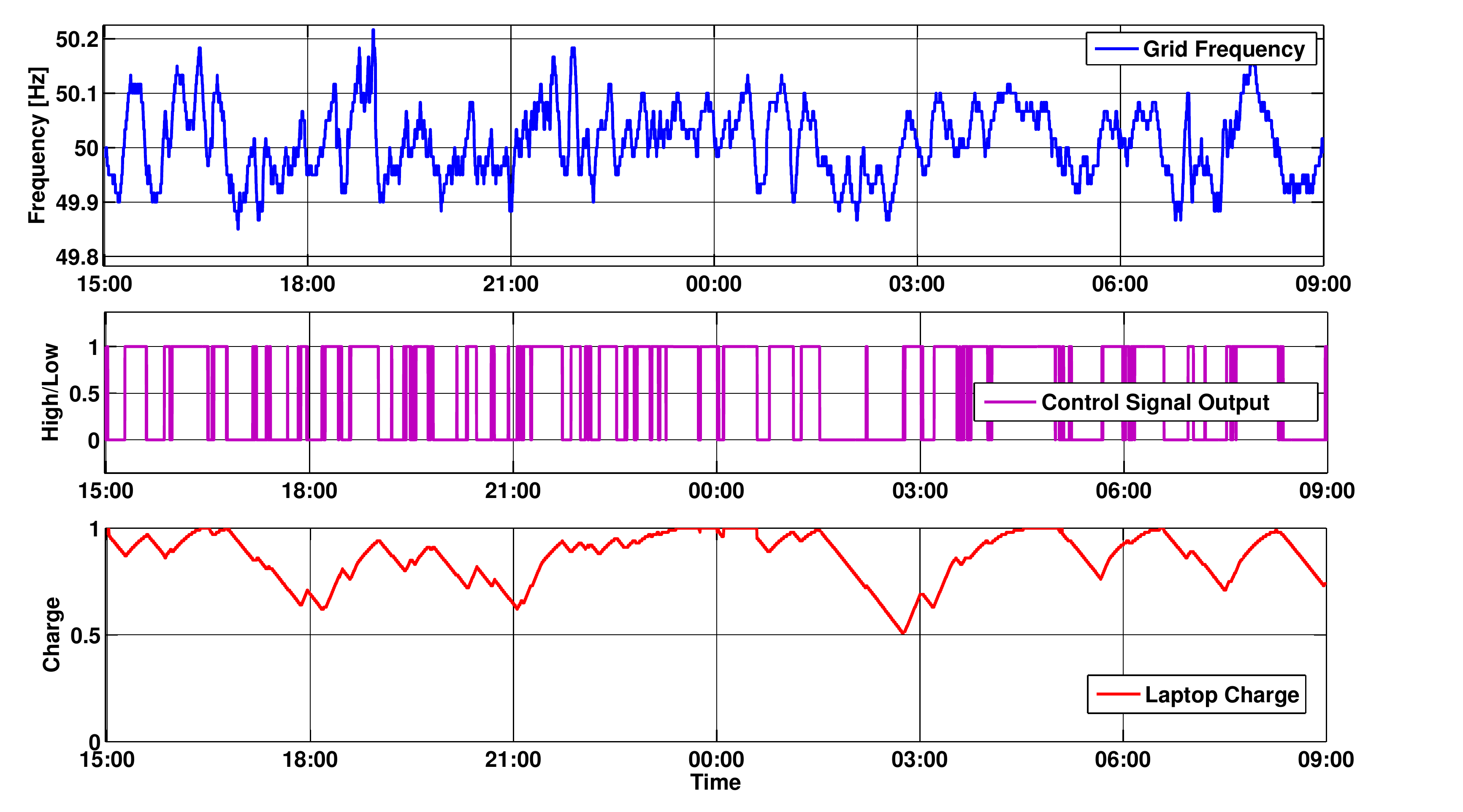}
\caption{Smart charging of a laptop using the setup shown in Figure \ref{fig:CDS_setup}. The upper plot 
shows the locally measured grid frequency for a period of 18 hours.
The center plot shows the output signal of the smart charging controller, 
whilst the bottom plot shows the charge status of the laptop, where 1.0 
and 0 correspond to 100\,\% and 0\,\% charge, respectively. It can clearly be seen that the 
battery charge decreases during periods when the grid frequency is below 50\,Hz and the laptop 
runs on its battery, while it recharges the battery or trickles whenever 
the grid frequency is above 50\,Hz.} \label{fig:dell18hours}
\end{figure*}

\section{Prototype of a smart charging laptop based on an external on-off
controller for the laptop power supply}
\label{sec:CDS}

For the first simplified prototype of a smart charging laptop we used a combination of 
simple analog electronics and a real time Linux digital control system \cite{CDS}, which 
is available in our lab. Figure \ref{fig:CDS_setup} shows a schematic of the 
setup. This  prototype measures whether the grid frequency is above a
certain threshold and if so it switches on the charging of the laptop.

An AC adaptor (transformer) is used to to obtain a pick off voltage of 5\,V AC.
This signal is first low pass filtered to prevent aliasing and then digitized with an analog-to-digital converter (ADC)
at a sampling frequency of 20\,kHz. Next we remove higher harmonic components as well 
as any unwanted DC components 
by filtering with a bandpass from 45--55\,Hz. The resulting signal is split into two paths in order to perform a differential 
measurement using two bandpass filters (Butterworth, 6th order), one tuned to be slightly below the threshold frequency
(50.0\,Hz in the setup shown in Figure \ref{fig:CDS_setup}) and the second one tuned to be 
slightly above the threshold frequency.  
If the actual grid frequency is above the threshold frequency, the amplitude after the bandpass stages
 will be larger in 
the right hand branch than in the left one. In order to get a DC measurement of the amplitude after the 
bandpass filter, we convert the signal into a symmetric rectangular voltage and take the absolute 
value. After low-passing, the signals from the two branches  are subtracted from each other 
and passed through a second low pass filter. A comparator is used to set the output of 
the digital controller to high if the signal after the subtracter is smaller than 0 (i.e. if the frequency
is above the threshold) and to low if the signal after the subtracter is greater than 0 
(i.e. if the frequency is below the threshold). Finally, the digital signal is output via a digital-to-analog converter (DAC)
and used to switch a relay, which can connect or disconnect the AC adaptor from the supply voltage.

Figure \ref{fig:dell18hours} shows the performance of this simple on-off prototype over a period of 18 hours and 
using a frequency threshold of 50.00\,Hz.
The top trace shows the time series of the grid frequency, varying between about 49.85 and 50.20\,Hz.
The output signal of the digital control system is shown in the center trace, with y-values of 1 representing 
active charging and y-values of 0 indicating that the charging is suspended. Finally the bottom trace 
shows the battery charge of the connected laptop (with charge level normalized between 0 and 1), a Dell Latitude 5520, which was connected to a power 
socket and in normal operation (playing music and keeping the screen lit continuously to simulate a typical  workload) during the whole measurement
time. When the 
grid frequency drops below 50\,Hz the laptop suspends the  charging process and runs on 
its battery and therefore the charge decreases. Whenever the frequency is above 50\,Hz
the charging is switched on again and the charge increases until the battery is fully charged. 
For this randomly chosen stretch of time the charge of the laptop varied between about 50\,\% and 100\,\%. 

\begin{figure}[h]
\centering
\includegraphics[width=\columnwidth]{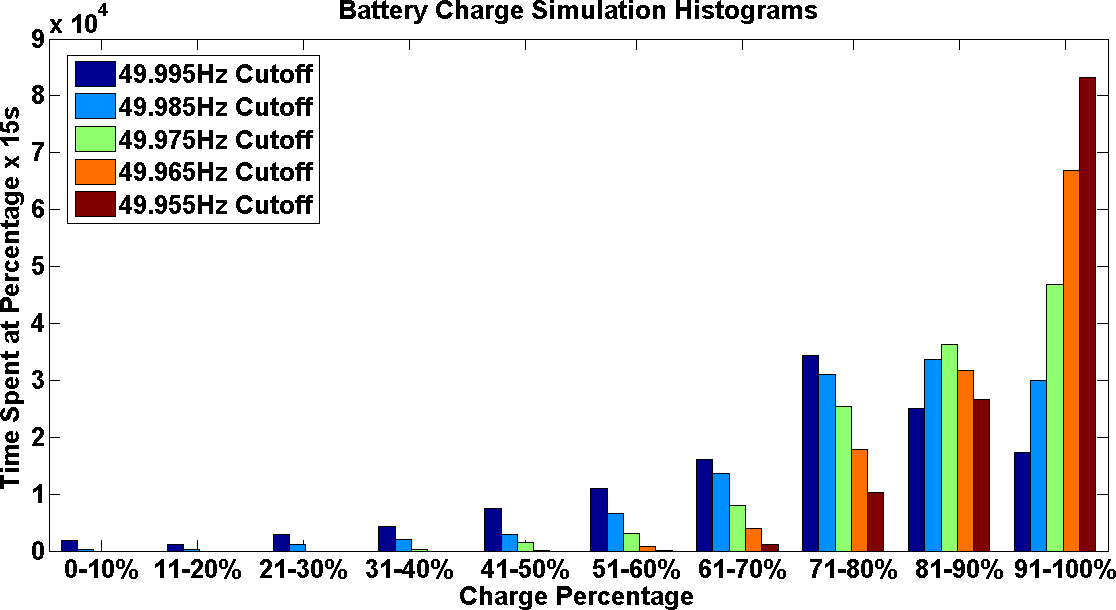}
\caption{Comparison of the charge status of our smart charging prototype connected to the Dell laptop (\emph{Laptop 1}) for
the application of different threshold frequencies. In order to guarantee comparability of the results, we developed 
a simulation of the battery charge that runs a virtual smart charging controller for different threshold frequencies 
over exactly the same grid frequency data set of 21 days duration. The lower the chosen frequency 
threshold, the higher the average charge of the laptop. While for threshold frequencies of 49.995 and 
49.985\,Hz Laptop 1 could occasionally run out of charge, for threshold frequencies of 49.975\,Hz and below
it would always keep at least a charge of larger than 30\,\%. Please note that these results are 
specific for Laptop 1 and should only be seen as an illustrating example. As shown in 
Appendix \ref{app:dellmac}, Laptop 2  would provide a significantly higher average 
battery charge and even for a threshold of 50.000\,Hz would not usually run out of charge.  }  \label{fig:simulation_histo}
\end{figure}

The charge status of the laptop obviously depends not only on the evolution of the grid frequency,
but also on the rates of charge and discharge. The slower the rate of discharge and the larger 
the capacity of the battery, the higher the average charge level of the battery. Similarly, the faster the 
rate of charge, the higher the average charge level. In addition to the Dell laptop 
(in the following referred to as \emph{Laptop 1}), we also tested 
the smart charging controller with a MacBook Pro$\textsuperscript{\textregistered}$
 laptop (in the following referred to as \emph{Laptop 2}),
 which yielded a higher average charge value
because of a higher ratio of rates of charge and discharge (see Figure
 \ref{fig:compare_dellmac} in Appendix \ref{app:dellmac}).
 In addition, the longer 
battery life of Laptop 2 made it possible for Laptop 2 to bridge even 
several-hour long stretches of the grid frequency
being below 50\,Hz, whereas the battery of Laptop 1 would  have been 
flat after about 2.5 hours (even when started from a fully charged battery).

One way to increase the average battery charge, especially for laptops such as Laptop 1 with a poor charge to discharge ratio or a degraded battery, is to reduce the frequency threshold to 
below 50\,Hz. Obviously this comes at the expense  of reduced mitigation of grid fluctuations, as the laptop
still charges even though the grid frequency has fallen slightly below 50\,Hz. However, in periods when the grid frequency 
is significantly below 50\,Hz, the smart charging laptop with a reduced threshold frequency still provides
the full level of mitigation.

Using real frequency data from the UK grid, we developed simulations of the battery status of our 
Laptop 1 for various frequency thresholds. The result for a 21 day long dataset is shown in 
Figure \ref{fig:simulation_histo}. While for threshold frequencies of 49.995 and 49.985\,Hz the 
Dell laptop still occasionally runs out of battery, lowering the threshold to 49.975\,Hz seems to guarantee 
that the laptop charge always stays above 40\,\%.

\section{Miniaturization of our hardware prototype using a microcontroller}
\label{sec:uC}

\begin{figure}[h]
\centering
\includegraphics[width=\columnwidth]{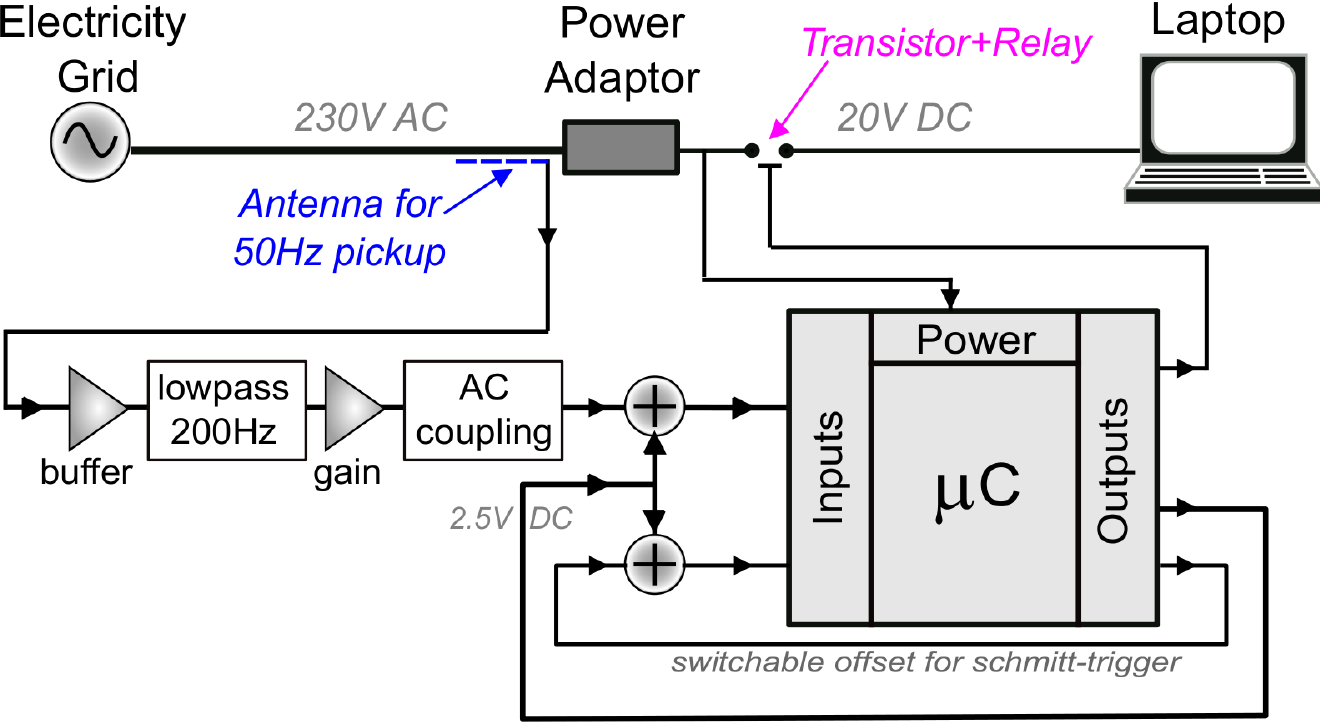}
\caption{Simplified schematic of miniaturized hardware prototype using a microcontroller.
The grid frequency is first picked up by a small antenna, then cleaned and amplified by analog electronics
and then measured with a microcontroller (see detailed description in the text). Depending on 
the actual grid frequency the mircrocontroller then gives out a control signal which is 
used to steer the charging process of the laptop. In order to guarantee the physical portability of 
the system, all electronic components and the microcontroller are powered from 
the laptop's power adaptor.} \label{fig:uC_setup}
\end{figure}

Since the hardware prototype  described in the previous 
section relied on a stationary digital control system, we then developed a miniaturized and portable version of a smart laptop 
charger. The simplified setup is shown  in Figure \ref{fig:uC_setup}.
Instead of the digital control system, in this prototype an Arduino$\textsuperscript{\textregistered}$
 \cite{Arduino}  microcontroller sits at the heart 
of the setup, performing the frequency measurement and controlling the charging process.

For the grid frequency measurement we used a short cable attached to the 
last few centimeters of the 230\,V power cable. The signal picked up was further 
processed using simple analog electronics. First the signal is buffered, then
passed through a low pass filter to remove any high-frequency noise, then it is amplified
to about 1\,V$_{\textrm{pp}}$--2\,V$_{\textrm{pp}}$ and finally AC coupled. As the inputs of the microcontroller 
are unipolar, we added an offset of 2.5V onto the signal and input this signal into
one ADC of the microcontroller and the offset without the signal into a second
input.

The actual frequency measurement in the microcontroller is performed in the time
domain \cite{freq_lib}. A comparator compares the two input signals and triggers 
an interrupt whenever the difference of the two inputs changes sign. The frequency
of the signal is then calculated by counting the clock cycles between consecutive interrupts.
In order to increase the robustness of the frequency measurements against noise, 
a Schmitt trigger \cite{Schmitt38} is implemented by using one of the microcontrollers
outputs to add a switchable offset onto the threshold voltage. 

\begin{figure}[h]
\centering
\includegraphics[width=\columnwidth]{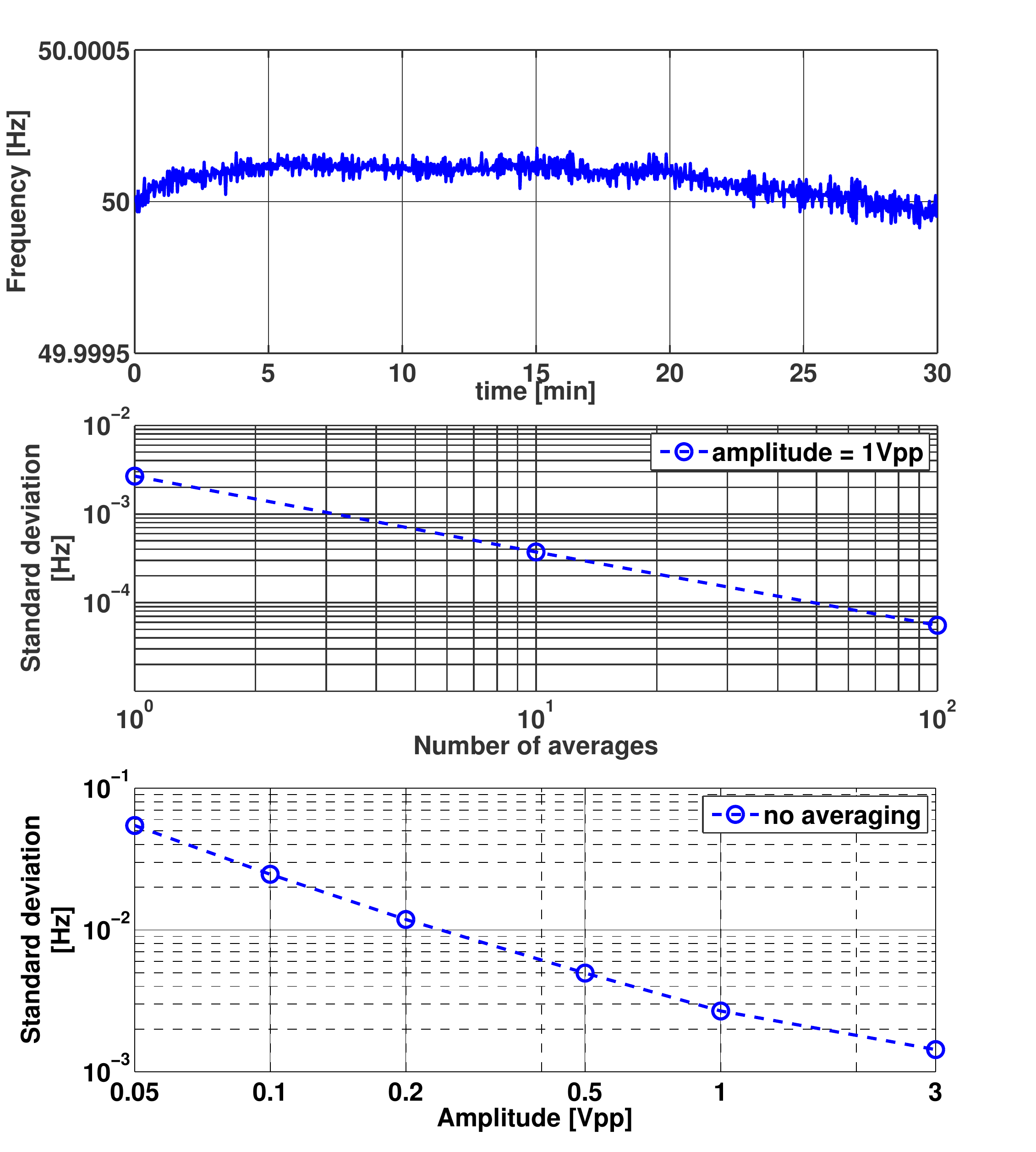}
\caption{Performance of the frequency measurement carried out with our microcontroller
setup whilst connected to a fixed reference frequency of exactly 50\,Hz.
Top plot: time series of a measurement with a signal amplitude of 1\,V$_{\textrm{pp}}$ and 
100 averages per measurement. Center plot: Dependence of the obtained frequency
resolution on the number of averages. Using 100 averages a frequency resolution 
of $5\times10^{-5}$\,Hz is achieved. Bottom plot: Dependence of the obtained frequency
resolution on the signal amplitude.  } \label{fig:uC_freq}
\end{figure}

In the microcontroller program one can set the number of  sign changing events over which the algorithm should 
average for a single frequency measurement. If no averaging is used a frequency 
measurement is performed every 20 milliseconds. As shown in the center plot of 
Figure \ref{fig:uC_freq}, without any averaging we obtain a frequency accuracy of about 2.5\,mHz. 
Using 100 averages gives a frequency measurement every 2 seconds with a frequency accuracy 
of about 0.05\,mHz, which should be more than sufficient for any smart charging application.
The microcontroller can, aside from the grid frequency measurement, be programmed with code 
implementing more sophisticated smart charging techniques, i.e. 
in our simple prototype we used a simple logic that switches the charging on if the grid frequency
is above one threshold and suspends the charging if it is below another threshold. These
two threshold frequencies do not necessarily need to be identical; in actual fact it is beneficial
to select slightly different thresholds for ``debouncing'', i.e. to avoid
 too frequent switching when the grid frequency is close to the 
thresholds. The charging control signal is then output by the microcontroller which 
controls via a transistor a relay in the circuit connecting the laptop and the power adaptor. In order to guarantee
physical portability and flexibility of our system, the microcontroller and all anolog components are powered directly 
from the laptop's power adaptor. Currently our smart laptop charger fits into a box of 
similar dimensions to the power adaptor itself. Using SMD technology, it will be  possible to
reduce the size to a few cubic centimetres.

Figure \ref{fig:uC_freq} shows the performance of the frequency accuracy of the microcontroller
setup. For these measurements the antenna was connected to a signal generator (Agilent 33120A) set to
exactly 50\,Hz. As discussed earlier we found that for a given signal amplitude the frequency resolution 
improves significantly with the number of averages. Moreover, the frequency resolution also improves 
with the signal amplitude (for a fixed number of averages), as shown in the lowest subplot. 
This is due to the steeper slope of the signal during the zero-crossing of the comparator. The 
top plot of  Figure \ref{fig:uC_freq} shows a timeseries of the frequency measurement using 100 averages 
and a signal amplitude of 1\,V$_{\textrm{pp}}$. We find the resolution to be much better than 0.1\,mHz and 
limited by a slow drift which is probably caused by a temperature driven perturbance of either the microcontroller's clock 
signal or the frequency stability of the signal generator itself \cite{Agilent}. 
However, the magnitude of this drift is too small to be of any concern for any potential application
of smart charging concepts. 

When comparing our two hardware prototypes  it has to be noted that the concept based
on the microcontroller has strong advantages over the realization using the setup described in Section 
\ref{sec:CDS}. The microcontroller
setup is much smaller and therefore portable. Also it requires much less and much cheaper hardware. Furthermore
it is of big advantage to the microcontroller setup that it has access to the absolute grid frequency as opposed to 
our implementation of a smart charger using the digital control system which is only able  to detect whether the grid 
frequency is above or below the set threshold.

\section{Software based realisation of smart charging}
\label{sec:soft}

As discussed in Section \ref{sec:principle} smart charging of portable electronic devices can also be realised 
entirely in software. Using a Levono Thinkpad T61
running the Ubuntu operating system we demonstrated the concept of a smart software charger which
supervises the laptop's battery management system (BMS). The software accesses the current grid
frequency via the Internet and reads in the 
battery charge status. The availability of the battery charge status allowed us to introduce a minimal charge level 
of the battery that is never surpassed as long as the laptop is connected to its power supply. Only above this minimal 
charge threshold the grid frequency is used to suspend or enable the charging of the battery.

\begin{figure}[h]
\centering
\includegraphics[width=\columnwidth]{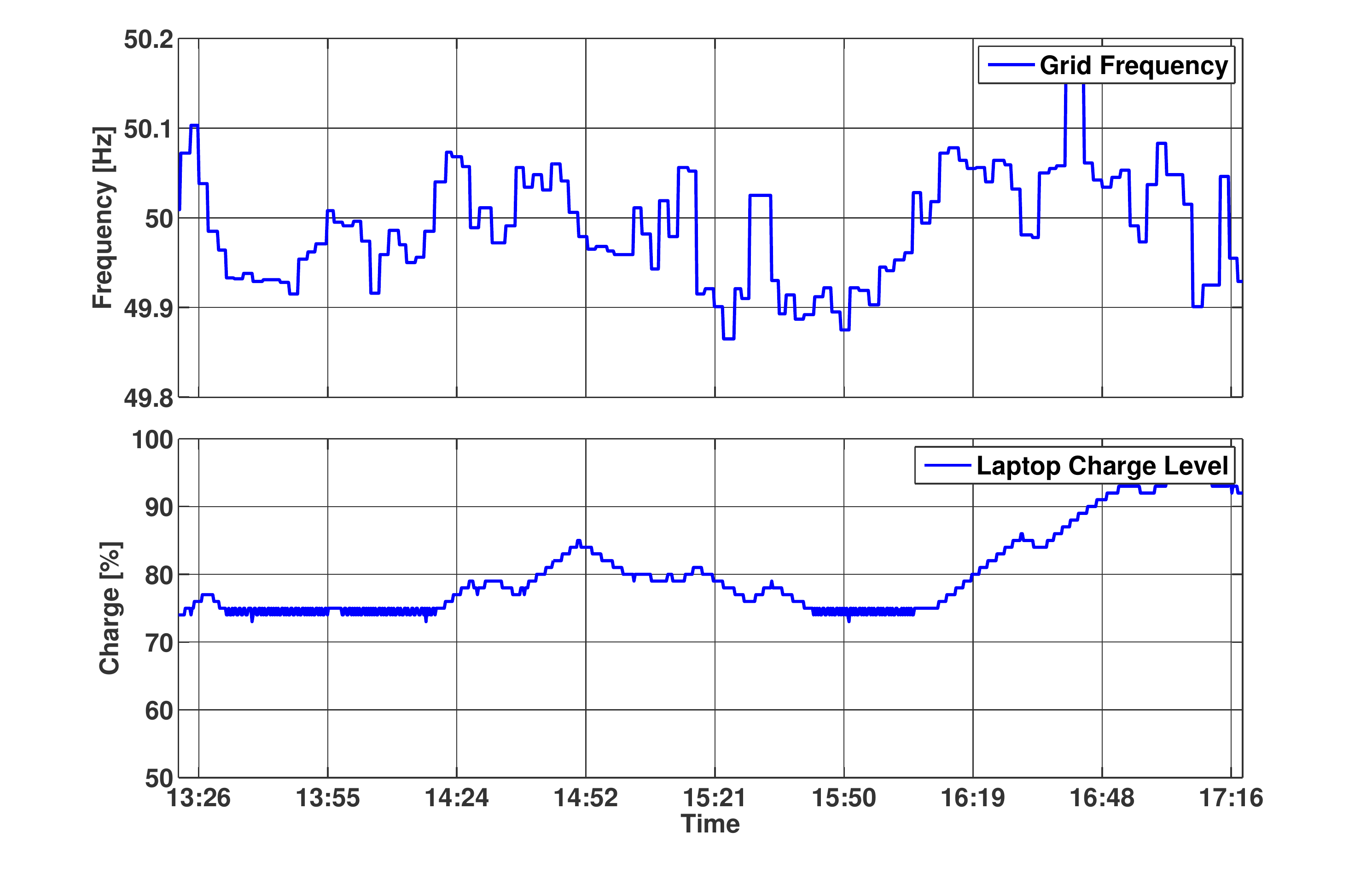}
\caption{Performance of a software based smart charger. Our software which supervises
the battery management system (BMS) controls the charging process depending on the grid frequency (received 
via the Internet) and the actual battery charge level. If the battery charge falls below a threshold defined in
the software (75\,\% in this example), then the battery will be charged, even when the grid frequency is 
below 50\,Hz.
 } \label{fig:soft1}
\end{figure}

\begin{figure}[h]
\centering
\includegraphics[width=\columnwidth]{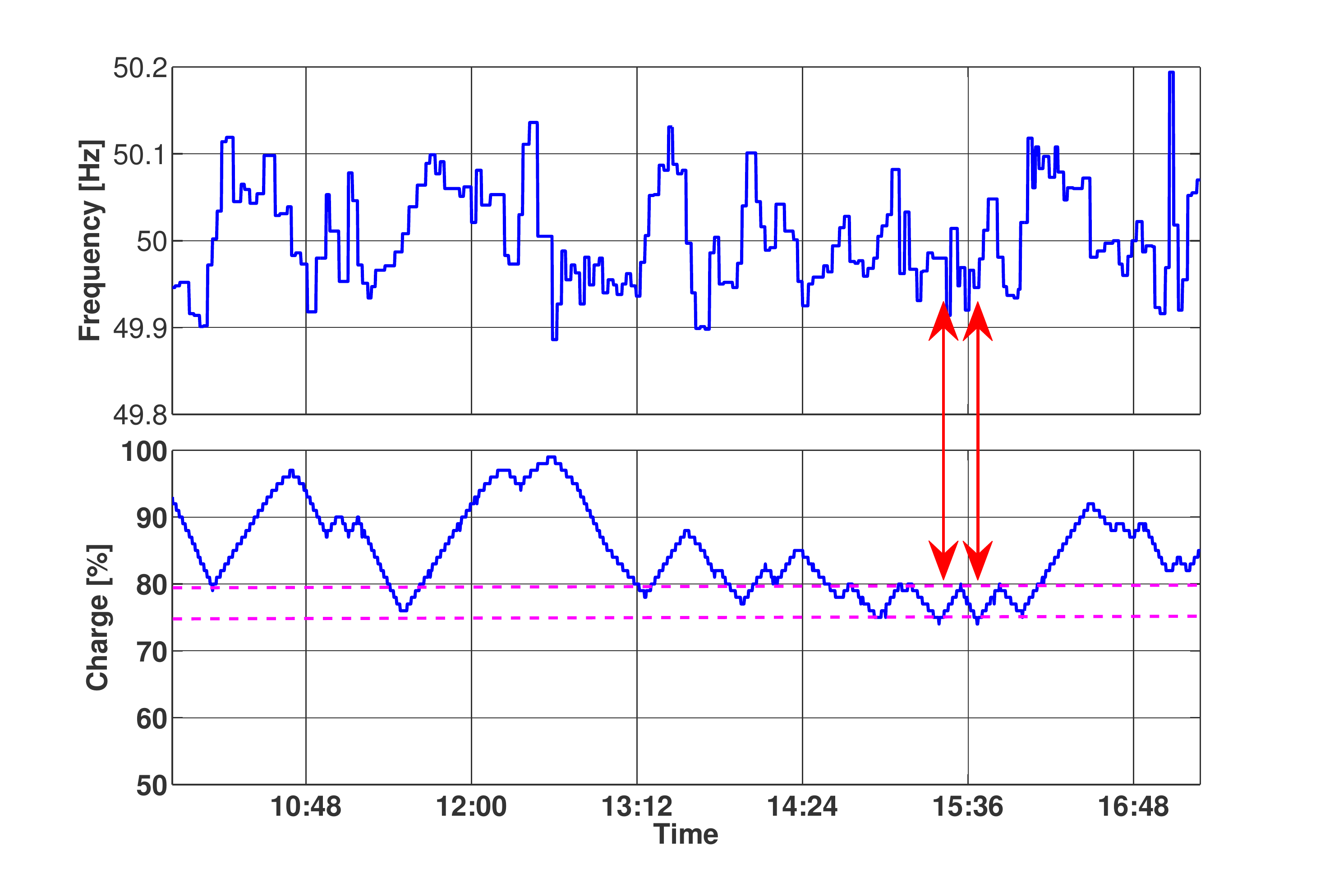}
\caption{Performance of a software based smart charger employing a two threshold method to avoid quick 
successions of switching on and off the charging process as it has been observed in Figure \ref{fig:soft1}. The 
two thresholds (pink dashed lines) are set to 75\,\% and 80\,\% in this example.
 } \label{fig:soft2}
\end{figure}

Figure \ref{fig:soft1} shows the performance of our software based charger using a minimal battery charge 
threshold of 75\,\%. In case the grid frequency is below 50\,Hz and the battery charge is at the minimal battery charge
threshold, this implementation causes the charging process to be suspended and resumed in quick
succession. 
A way to avoid this very frequent switching of the charging we implemented in a further step a two-threshold concept as 
displayed in Figure \ref{fig:soft2}. We used a lower threshold of 75\,\% and a higher threshold of 80\,\%. When the battery 
charge level drops below 75\,\% the battery is charged independently of the grid frequency. Only when the battery charge 
reaches the upper threshold at 80\,\% will the charging controller start to take into account the grid  frequency again. As a result,
during the  two times indicated by the vertical arrows the laptop battery is charged even though the grid frequency is 
below 50\,Hz. However, in contrast to Figure \ref{fig:soft1} employing the double threshold method, the previously observed 
continuous switching on and off of the battery charging is avoided.  

The concept described in this section, which does not only take the grid frequency into account, but includes 
the actual battery charge level and steers the charging process via supervising the BMS, advances smart charging 
of portable electric devices beyond the external on-off controller 
for the power supply. 

One clear disadvantage of the software based smart charger described here is the fact that it requires
the laptop to have access to the Internet and therefore the laptop needs to be running when the smart charging 
modus should be active.

\section{Future Developments}
\label{sec:future}

The two hardware prototypes for a simple on-off  smart charging controller as discussed in  sections \ref{sec:CDS} and \ref{sec:uC}
together with the software based smart charging concept described in the previous section
constitute a first but nonetheless import step towards the practical realization of our smart charging concept.

In order to utilise the full potential of the concepts described
in Section \ref{sec:principle} it will be essential that the smart controller also has access 
to habits of the actual user.

Having access to the  grid frequency, the battery charge level  and the individual user profile/preferences
would allow the implementation of more complex control algorithms which could 
include individual or combinations of the following features:
\begin{itemize}
\item A certain available charge status at particular user-defined times, such as to have the battery always hold at least 80\,\% charge as of 08:00 on weekdays.
\item Instead of using a `hard' frequency threshold, a weighting factor could be introduced 
that takes into account not only of whether the grid frequency is above or a below the threshold, but also 
by how much the frequency is above or below the threshold. Assuming that the deviation of the 
grid frequency from 50\,Hz is proportional to the supply-to-demand ratio, it will be possible 
to trade off the urgency of required demand suspension versus the user's urgency to charge their electronic devices on a particular occasion.
\item Finally, it is also possibly desirable to investigate smart charging algorithms based on advanced 
control concepts, such as adaptive filters (for instance Wiener filters \cite{Wiener}) or self-learning systems,
such as neural networks 
that can be optimised for the user's particular habits as well as for the characteristic demand evolution on 
daily, weekly and seasonal time scales.
\end{itemize}
  
At the moment it is not clear whether it will be best to have the complete smart charging algorithm 
entirely implemented into individual portable electronic devices, or whether it will be more efficient
to have certain parts of the algorithm which are common for all users inside the electricity grid 
(such as the measurement of the grid frequency) performed centrally with the resulting information
 and/or steering signals being sent to the individual charging devices.  
 
 Furthermore, future efforts will include a detailed analysis of the large-scale effect and benefits of 
 our concept onto the electricity grid when employed in millions of devices. Also we plan to investigate 
 the size and cost requirements of a hardware controller based on a purpose-built microchip which could
 be integrated directly into the power supply of portable electronic devices
 by manufacturers.

\section{Conclusion}
\label{sec:con}
Demand side control techniques and smart grid concepts are currently under consideration
or are in the process of being implemented for a wide range of applications. In this article we have reported on our efforts 
to extend these techniques to applications for portable electronic devices featuring internal 
energy storage, such as laptop computers, mobile phones and tablet computers. We believe that 
the concepts we have described in this article could be applied in the medium term future to the majority
of the portable electronic devices produced. This would not only contribute to the stabilisation 
of national electricity grids, helping to mitigate the electricity production fluctuations from
renewable energy sources and helping to avoid the building of costly standby or backup power plants, 
but would also allow the end user to save money, by predominantly using electricity for charging their electronic devices in
 periods when it is cheap.
 
 We have developed hardware and software based prototypes of a smart charging concepts for laptop computers. To the best 
 of the authors' knowledge this was the world's first demonstration of demand control techniques 
 applied to laptop computers. The hardware prototypes continuously perform a local measurement of the 
 grid frequency and switch on or suspend charging, depending on whether the grid frequency is above or below a pre-set threshold, respectively. While the first prototype made use of a stationary
 digital control system and applied digital filtering techniques to analyse the grid frequency, the 
 second prototype is based on a microcontroller to reduce the size of the device and the
  frequency measurement is performed by counting sign changes of the signal.
 
 Our tests have shown that the smart charging concepts for portable electronic devices we described
 work entirely satisfactory. Moreover we found that the actual performance strongly depends on the certain
 properties of the underlying electronic devices, such as the overall battery life time and the ratio of the 
 rates of charge and discharge. For systems with low battery lifetime and a low ratio of rates of charge 
 and discharge, it can be preferable to use a frequency threshold of a few tens of mHz below 50\,Hz to make sure
 that the electronic devices never run out of charge. 
 
In addition we also developed a software based smart charging algorithm that steers the charging 
 process of a laptop by supervising the battery management system and switching the charging on and off
 depending on the grid frequency as well as the user-defined battery charge level thresholds.
 
 Future developments include investigations into how best to combine the hardware and 
 software approaches. We also intend to test more complex charging controllers which include the personal 
 user habits as an additional input to the smart charging algorithms. Moreover, the application of advanced charging algorithms,
 based on for instance adaptive filters or neural networks, will be investigated.

\appendices
\section{Charging and discharging characteristics of the laptop computers we tested}
\label{app:dellmac}

As mentioned in Section \ref{sec:CDS} the overall performance of any smart charging technique
for an electronic device strongly depends on its battery capacity and its rates of charge and discharge. 
We define the ratio of charge and discharge rates to be:
\begin{equation}
R = \frac{\textrm{Rate of Charge}}{\textrm{Rate of Discharge}}.
\end{equation}

\begin{figure}[h]
\centering
\includegraphics[width=\columnwidth]{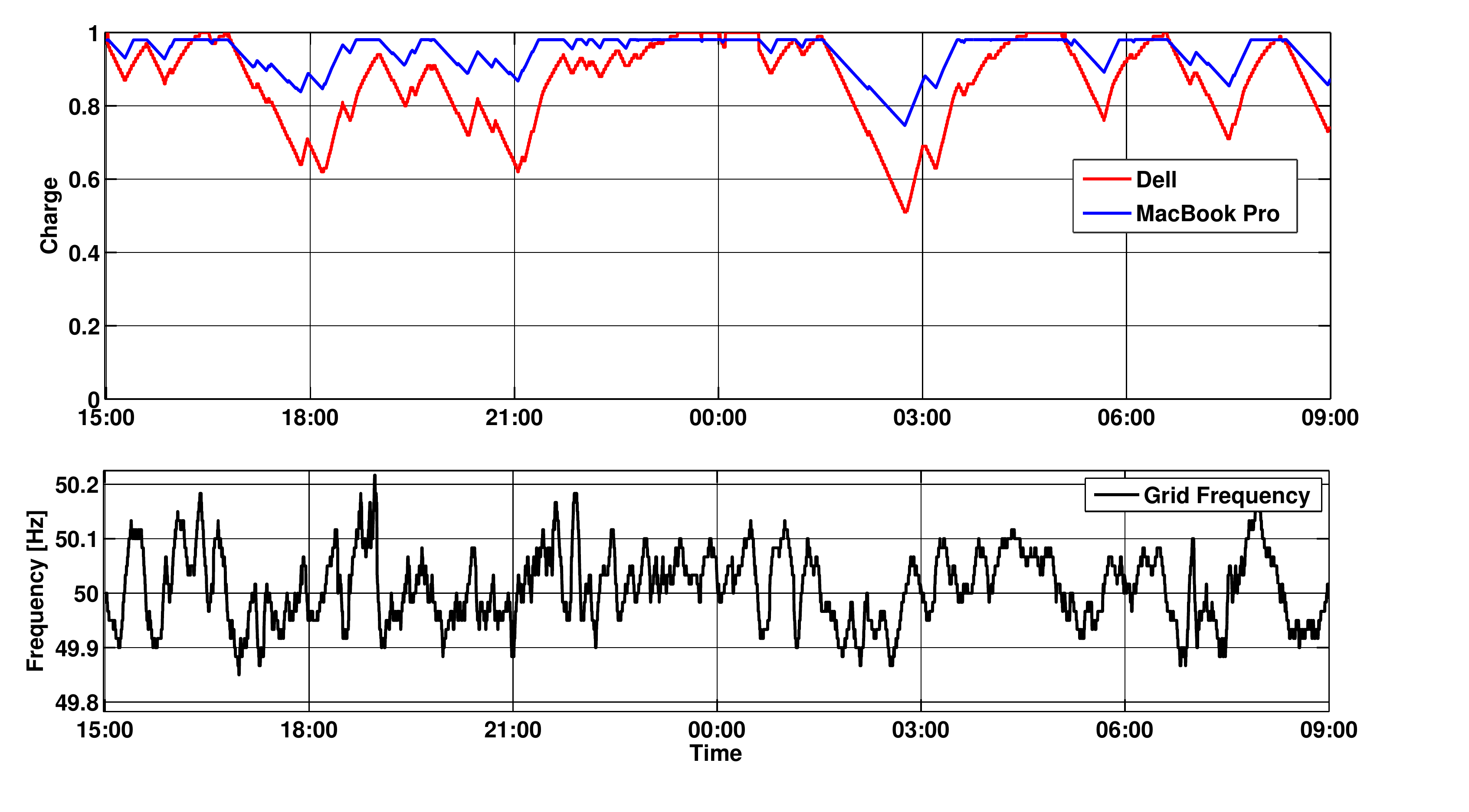}
\caption{Comparison of the measured charge status of Laptop 1 and the simulated 
charge status of Laptop 2 for the same time stretch as displayed in 
Figure \ref{fig:dell18hours}. For both laptops a frequency threshold of exactly 50\,Hz is used. 
Due to its higher charge to discharge rate and its 
significantly larger battery lifetime, Laptop 2 would provide a significantly 
higher average battery charge than Laptop 1 when operated in a smart charging 
mode.
} \label{fig:compare_dellmac}
\end{figure}

The charge and discharge rates as well as their ratio for the laptops tested
can been found in Table \ref{tab:chargeanddischargerates}. 
\begin{table}[h]
\begin{centering}
	\begin{tabular}{| l | c | c | c |}
	\hline
	\textbf{Laptop} & \textbf{$R_{\textrm{Charge}}$} & \textbf{$R_{\textrm{Discharge}}$} & $R$ \\
	\hline
	Laptop 1 (Dell E5520) & 0.0213 & 0.0115 & 1.8522 \\
	\hline
	Laptop 2 (MacBook Pro$\textsuperscript{\textregistered}$) & 0.0128 & 0.0056 & 2.2857 \\
	\hline
	\end{tabular}
	\caption{A list of the rates (percentage points per second) of charge, $R_{\textrm{Charge}}$, and discharge, $R_{\textrm{Discharge}}$, and the ratio between the two.}
\label{tab:chargeanddischargerates}
\end{centering}
\end{table}
The discharge rate indicates how long it would take the laptop to go from fully charged
down to a flat battery during normal operation. This duration we define as \emph{battery life time}, and it turns out to be only about 2.5 hours 
for Laptop 1, whereas Laptop 2 has a battery life time of about 5 hours. Therefore, when used in combination 
with a smart charging controller, Laptop 2 shows better performance during extended periods (of the order hours) 
with a grid frequency below the charging threshold. Moreover, due also to its higher value for the ratio
 $R$, Laptop 2 regains its 
battery charge about twice as quickly as Laptop 1 during periods when the grid frequency is above the threshold.  
The result of these performance differences between Laptop 1 and Laptop 2 can be seen in Figure
 \ref{fig:compare_dellmac} which indicates the battery charge evolution of both laptops for the same period of 
 time. While for this particular stretch of time the charge of Laptop 1 decrease to values as low as 50\,\%, the 
 battery charge of Laptop 2 always stays above 75\,\%.

%

\section*{Acknowledgment}

The authors would like to thank Kenneth Strain, Andreas Weidner and Mark Kille for fruitful 
discussions. We are also grateful for financial support from the John Roberston Bequest 
as well as the College of Science and Engineering at the University of Glasgow. 

\ifCLASSOPTIONcaptionsoff
  \newpage
\fi

%

\begin{IEEEbiographynophoto}{Stefan Hild}
received a physics degree (Dipl. Phys) from the University of Hannover, Germany, in 2003.
In 2007 he received his PhD (Dr. rer. nat) from the Max-Planck Institute for Gravitational Physics
(Albert-Einstein Institute) and the Leibniz University of Hannover for his  work on the commissioning 
and enhancement of the British/German laser-interferometric gravitational wave detector GEO\,600. 
After 2 years of postdoctoral research in Astrophysics and Space Research group of the University 
of Birmingham, UK, he obtained a faculty position at the University of Glasgow. Stefan Hild has a 
keen interest high-precision interferometry, quantum-non-demolition measurements as well as 
research into renewable energy and smart grid applications. He is the chair of the Sensing and 
Control Group of the GEO\,600 collaboration.  
\end{IEEEbiographynophoto}
\begin{IEEEbiographynophoto}{Sean Leavey}
 received the MSci degree in physics from the University of 
Glasgow, United Kingdom, in 2012. He is currently undertaking a 
PhD degree at the Institute for Gravitational Research in the University of Glasgow, 
working on the instrumentation for ground-based, laser-interferometric  
gravitational wave detectors. Additionally, his research interests 
include the investigation of solutions for the problems facing future 
electricity grids.
\end{IEEEbiographynophoto}
\begin{IEEEbiographynophoto}{Christian Gr\"af}
received his Dipl.~Phys.~degree in physics from the Leibniz University of Hannover, Germany, in 2008.
His current work focuses on optical design aspects of a prototype laser interferometer which aims
for reaching and later surpassing the standard quantum limit for its 100 gram test masses. 
Besides laser interferometry his research interests include electronics and numerical optics.
\end{IEEEbiographynophoto}
\begin{IEEEbiographynophoto}{Borja Sorazu}
 graduated in physics (specializing in electronic and control engineering) from the University 
of the Basque Country (UPV), Bizkaia, Spain, in 2000, and then he graduated in electronic engineering 
from the same university in 2001. Subsequently he received the Ph.D. degree in electronic and electrical 
engineering (specializing in optical fibre sensors and laser ultrasonics) from the University of Strathclyde, 
Glasgow, U.K., in 2006. During his Ph.D. he was involved in the use of optical techniques for structural
 examination of mechanical systems and material evaluation, including signal generation and 
 acquisition and data interpretation.
He has since joined the Institute for Gravitational Research (IGR) at the University of Glasgow, U.K., as 
a Research Associate, where his current research interest focus on several aspects of advanced interferometry 
for application to ground based gravitational wave detectors, and the detector characterization of GEO 600. 
\end{IEEEbiographynophoto}




\end{document}